\documentclass[conference]{IEEEtran}
\IEEEoverridecommandlockouts

\usepackage{cite}
\usepackage{amsmath,amssymb,amsfonts}
\usepackage{algorithmic}
\usepackage{graphicx}
\newcommand{\mytablesize}{\scriptsize}
\usepackage{textcomp}
\usepackage{xcolor}
\usepackage{booktabs}
\usepackage{makecell}
\usepackage{xcolor}
\usepackage{xurl}
\usepackage{tabularx}
\usepackage{subcaption}
\usepackage{url}
\usepackage[most]{tcolorbox}

\def\BibTeX{{\rm B\kern-.05em{\sc i\kern-.025em b}\kern-.08em
    T\kern-.1667em\lower.7ex\hbox{E}\kern-.125emX}}
\begin{document}

\newcolumntype{Y}{>{\centering\arraybackslash}X}
\newcolumntype{R}{>{\raggedleft\arraybackslash}X}
\newcolumntype{L}{>{\raggedright\arraybackslash}X}

\title{Beyond the Wrapper: Identifying Artifact Reliance in Static Malware Classifiers using TRUSTEE
}

\author{
\IEEEauthorblockN{Riyazuddin Mohammed}
\IEEEauthorblockA{\textit{School of Informatics, Computing, and Cyber Systems} \\
\textit{Northern Arizona University}\\
Flagstaff, AZ, USA \\
rm2883@nau.edu}
\and
\IEEEauthorblockN{Lan Zhang}
\IEEEauthorblockA{\textit{School of Informatics, Computing, and Cyber Systems} \\
\textit{Northern Arizona University}\\
Flagstaff, AZ, USA \\
Lan.Zhang@nau.edu}
}

\maketitle

\begin{abstract}
Modern cybersecurity relies heavily on static machine-learning-based malware classifiers. However, transformations such as packing and other non-semantic modifications applied to executable files limit their reliability.  Malware classifiers often learn these unnecessary artifacts rather than the true binary behavior because of the high association between maliciousness and packing. Moreover, these malware classifiers are black boxes, making it difficult to understand what they learn. To address this issue, we proposed a two-part framework using the post-hoc interpretability XAI tool TRUSTEE, followed by a manual analysis of the top features. We conducted several controlled experiments by varying the dataset composition ratios to understand their impact on the results. The top-ranked features across all experiments, identified by TRUSTEE, were predominantly packing artifacts, portable executable(PE) metadata, and $n$-grams at the string level, rather than malicious semantics. These results suggest that these malware classifiers are highly sensitive to dataset composition and can misinterpret packing as malicious behavior. Our proposed framework allows for the reproducible diagnosis of such biases and forms a guideline for building more robust and semantically meaningful malware detection models.

\end{abstract}


\section{Introduction}

The use of signature-based methods for classifying executables as benign or malicious has become increasingly challenging due to the evolution of malware families and the increasing volume of data. These challenges are addressed by modern malware detection techniques that use machine-learning models to classify executables. ML models are widely employed in the field of cybersecurity for malware detection~\cite{saxe2015deep}~\cite{gibert2020rise}~\cite{rathore2018malware}. These ML malware classifiers are trained on the features extracted by applying static analysis to executable files to classify any executable file as benign or malicious. 
One such factor is the packing applied to the executable file to achieve code protection, obfuscation to resist reverse engineering, and faster and safer distribution. However, many malware authors often use packing as camouflage to evade detection by these static malware classifiers~\cite{5633410}, which creates a strong correlation of malware with packing. This misconception also arises from the imbalance in real-world malware classification datasets, where packed malicious samples significantly outnumber packed benign samples~\cite{almajed2025imbalance}. Recent works in the literature have demonstrated that these machine learning (ML) malware classifiers depend on packing artifacts instead of meaningful behavioral semantics~\cite{pendlebury2019tesseract}, thus raising several concerns about the reliability and robustness of these static ML classifiers~\cite{gibert2025assessing}.
Thus, raising the question \textit{``Do these models actually learn malicious binary behavior, or are they primarily relying on some other artifacts that correlate with maliciousness ?"}~\cite{guidotti2018survey}~\cite{alvarez2018robustness}~\cite{perasso2025empirical}

Some recent studies of Explainable AI (XAI) in security also showed that models rely on unintended features for classification rather than on the features that experts in the field consider important. However, the classifiers used were white-box, and the metrics used were coarse metrics, such as true positives and true negatives, rather than interpreting the features to understand what malware classifier actually learned~\cite{sharma2018detection}~\cite{aghakhani2020malware}. Misclassifications in security-critical environments poses higher security risks which increases the need for robust and interpretable models.

The shift in the use of ML classifiers in modern malware detection, along with the conclusions drawn from recent studies, motivated us to determine whether black-box malware classifiers rely on unintended features for classification. This can be achieved by carefully analyzing the top features prioritized by existing XAI tools, without which the importance scores for those features remain opaque~\cite{guidotti2018survey}.

Despite all the advancements in malware detection and interpretability, the following challenges need to be addressed to solve the issues discussed in previous studies.

\begin{itemize}
    \item[\textbf{C1:}] \textbf{Unstable Interpretability Outputs:} Lack of stable outputs from XAI tools like LIME~\cite{ribeiro2016should}, SHAP~\cite{lundberg2017unified}, and surrogate trees because of sampling in the dataset~\cite{alvarez2018robustness}.
    \item[\textbf{C2:}] \textbf{Identification of key features:} The malware classifier model learning packing artifacts or PE file metadata instead of actual true binary~\cite{perasso2025empirical}.
    \item[\textbf{C3:}] \textbf{Dataset Composition Bias:} Even minor imbalances between the Unpacked Benign (UB), Packed Benign (PB), Unpacked Malware (UM), and Packed Malware (PM) categories can mislead classifier~\cite{pendlebury2019tesseract}.
\end{itemize}

To address challenges C1 and C2, we propose a two-part framework that combines:

(a) TRUSTEE~\cite{jacobs2022ai} proposed by Jacobs et al. is an application of models trained on minimally overlapping datasets for multiple iterations to make the XAI output stable for feature interpretation.

(b) Manual byte-level analysis of PE files to understand the features ranked as most important by TRUSTEE to classify them as true binary behavior or unnecessary artifacts.

To address C3, we trained independent models on different datasets with unique data composition ratios of UB, PB, UM, and PM samples to identify the dataset composition bias in the malware classification model. This enables us to understand the classifier's decisions based on the dataset composition after analyzing the top features.

Our experiments revealed that over 85-90\% of the features marked as important by TRUSTEE were not malicious binary behavior-related. Instead, the classifier relied on artifacts related to packing or PE metadata residues, or string-level $n$-grams, with less than 10–15\% reflecting true malicious semantics. Additionally, we investigate whether these learned feature dependencies remain consistent across different models and datasets. Unlike prior work that qualitatively identifies artifact reliance~\cite{perasso2025empirical}, we provide a controlled, reproducible framework that systematically isolates dataset composition effects and demonstrates how models adapt to alternative artifact signals under distribution shifts.

The contributions of this study are as follows:
\begin{enumerate}
    \item We formally investigate whether static malware classifiers learn malicious semantics or rely on packing artifacts.
    \item We propose a reproducible framework combining controlled dataset construction, repeated TRUSTEE-based feature extraction, and systematic manual feature interpretation.
    \item We demonstrate the effect of data composition ratios by varying the UB, PB, UM, PM numbers on the classifiers by using the proposed framework for building more robust classifiers.
    \item We demonstrate the basis for the malware classifier model’s decisions with evidence, offering guidance for building more reliable malware-detection models in the future.
\end{enumerate}

\section{Background}

\subsection{Machine Learning for Malware Detection}\label{AA}
Traditional signature-based methods often struggle with large volumes of software data for classification. In addition, as malware evolves over time, traditional signature-based methods struggle to deal with new malware families. To address these challenges, researchers and industry practitioners employ modern malware detection systems~\cite{gibert2020rise}. These systems train classifiers on existing datasets to distinguish software as benign or malicious by leveraging the statistical patterns learned from the data~\cite{5633410}.   

Many machine learning architectures, such as random forests and decision trees, have been deployed as malware classifiers because of their high interpretability. However, in recent days, the use of deep neural networks for malware classification has grown significantly because of their ability to process raw bytes and embeddings of binary content. The ability to generalize vast families of malware and the high accuracy performance offered by these deep neural networks are the major reasons for their acceptance by researchers~\cite{saxe2015deep}~\cite{gibert2020rise}~\cite{rathore2018malware}. The key assumption is that these models learn the characteristic structures and statistical signatures of malicious binaries that can be captured from data, thereby enabling automatic identification at scale.

Despite these advantages, the internal decision-making of modern malware classifiers is hidden, as they are black-box models. This raises concerns regarding their application in the field of security, which is a high-risk field. The trust needed by security domain experts to deploy ML-based malware classifiers drives the motivation to understand the internal decision-making performed by the malware classifier to classify any software as benign or malicious.  

\subsection{Static Analysis and Feature Extraction}
Static analysis is the process of inspecting executables without executing the program for malware detection. Structural and semantic information present in a Windows Portable Executable (PE) is extracted for the static analysis. The features extracted from this information were derived from the headers, metadata, and contents of the PE file. A family of nine static analysis features is extracted for the malware classifier to be trained, which includes (1) PE headers, (2) PE sections, (3) DLL sections, (4) API imports, (5) Rich Header, (6) Byte $n$-grams, (7) Opcode $n$-grams, (8) strings, and (9) file generic. These features collectively create a rich and high dimensional representation of the executable~\cite{aghakhani2020malware}. 

Recent works in the literature have shown that static features do not always represent the program behavior. Many PE metadata fields reflect compiler implementation details, build tools, or development environment characteristics rather than malicious intentions~\cite{rezaei2021pe}. Many structural patterns from packing methods and library dependencies are captured using byte-level and opcode $n$-grams. Consequently, the malware classifier may be distracted by these features and fail to focus on the features that capture actual program behavior~\cite{lyda2007using}~\cite{webster2017finding}~\cite{rezaei2021pe}.

\subsection{Packing and Its Impact on ML Classifiers}
One of the transformations applied to PE file is packing, which is applied to executables and has the greatest influence on malware detection. Packing is generally performed to change the structure of the binary while preserving the functionality to  achieve compression, encryption, and obfuscation of the executable content and metadata, thus reducing the size of the executable file. Packers such as Themida, UPX, and VMProtect often alter section entropy, which is a measure of randomness, alter the import tables~\cite{5633410}, change the PE sections, and may even rewrite or remove metadata such as the Rich Header. Thus, packing enables code size reduction, protection of proprietary logic, making the executable tamper-resistant to some extent, and faster distribution~\cite{webster2017finding}~\cite{aghakhani2020malware}.

All the above benefits of the packing process make it favorable for software developers, but it also encourages malware authors to use packing for malware. Malware authors often employ packing to hinder reverse engineering, conceal malicious payloads, disrupt disassembly, and limit the visibility. Static tools often find it difficult to identify malicious parts~\cite{aghakhani2020malware} because of the high entropy in packed executables caused by the encryption or compression of sections in the binary file and the high associativity of maliciousness with packing~\cite{almajed2025imbalance}.

Packing affects modern malware detection because of the alterations made to executable files. Real-world malware datasets used for training machine learning models are biased, with substantially higher malicious samples being packed~\cite{almajed2025imbalance}. All these factors increase the association of malware with the packing process, thus making the malware classifier model learn unnecessary relations rather than the true binary behavior of the binary. This phenomenon limits the generalization capability of static machine learning detectors and underscores the need for precise interpretability tools to understand the true basis of classifier decisions.

\subsection{Explainable AI for Malware Classification}

The growth in the use of malware detection machine learning models in the field of security has made it necessary to understand their behavior. Well-known Explainable AI (XAI) tools, such as SHAP and LIME,  analyze the predictions made by the model by determining the contributions of individual features for any decision made by the model. This can also be achieved by extracting interpretable surrogate models that represent complex decision boundaries~\cite{jacobs2022ai}. However, these approaches have several limitations in improving the interpretability of black-box models, particularly due to the large size of the datasets. Moreover, the attributions generated by most of these XAI tools are unstable as the variations are high depending on the sampling, small changes in the input or model initialization which raises concerns regarding the reliability of these methods~\cite{alvarez2018robustness}~\cite{jacobs2022ai}. The low-level analysis performed on features such as $n$-grams or raw byte sequences results in an insufficient explanation of the features attributed as being important. Prior interpretability research often provides results only at the level of coarse feature families, such as API imports, strings, or header groups, rather than examining the individual features that trigger classifier predictions~\cite{aghakhani2020malware}.

The TRUSTEE framework generates interpretable, low-complexity, and high-fidelity decision trees that approximate the internal behavior of a black-box model~\cite{jacobs2022ai}. Given a black-box model along with the dataset on which it is trained, the TRUSTEE can produce interpretable decision trees and trust reports. Although TRUSTEE claims that it can produce a stable output for small changes in the input, we ensured stable output generation in our proposed framework which is detailed in Section~\ref{met1}. This makes TRUSTEE an appropriate choice as our XAI tool, as it is also well-suited in the domain of malware detection to understand if our malware classifier is relying on the true behavior signals or not to evaluate the robustness or trustworthiness of the classifiers.

\section{Methodology}

\begin{figure*}[t]
    \centering
    \includegraphics[width=0.7\textwidth]{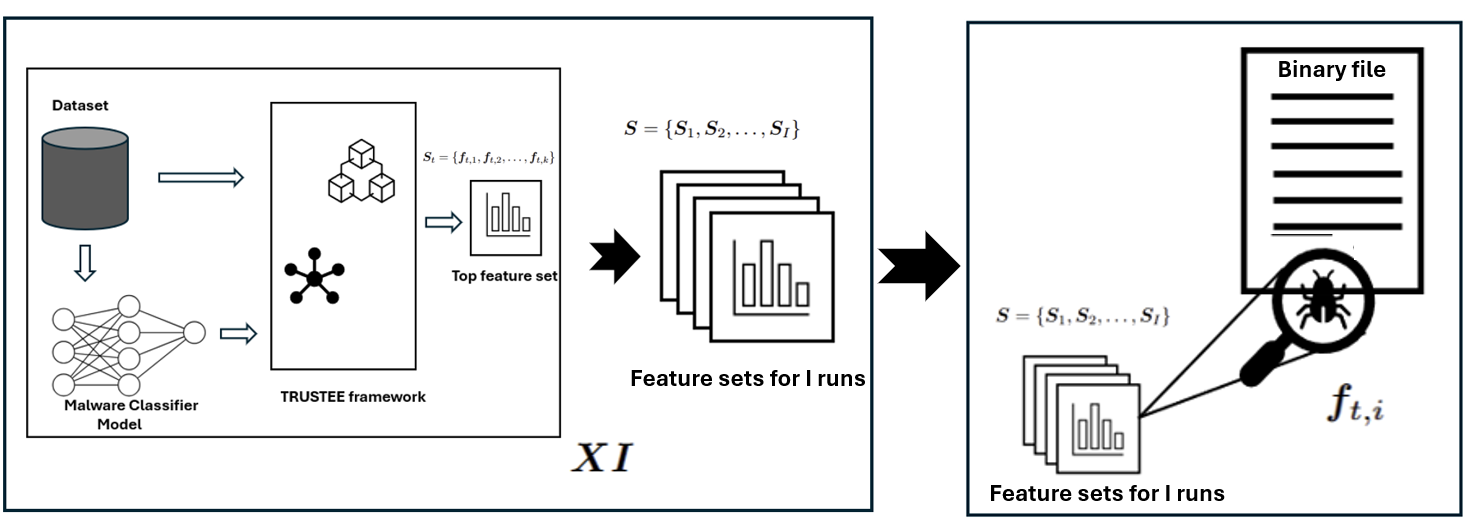}
    \caption{Overall pipeline of the proposed methodology.}
    \label{fig:pipeline}
\end{figure*}

To address the research question, \emph{Is the malware classifier learning the actual binary behavior or the packing behavior when classifying a sample as benign or malicious?}  \\we propose a trustworthy framework as illustrated in Figure~\ref{fig:pipeline}.

The proposed framework consists of two parts. They are, \textbf{1. Feature Extraction:} Extract the top features that the malware classifier relies on using the TRUSTEE framework.
\textbf{2. Feature Analysis:} Analyze the top features from feature extraction by performing a manual byte-level analysis on the PE files to determine whether the top features are binary-related.
\subsection{Extraction of Top Features}\label{met1}
To train the malware classifier $\pi$, we used a labeled dataset $D$ with benign and malicious samples. As we are concerned with the packing behavior, we also considered whether a sample is packed or not, which divides the samples into four categories: Unpacked Benign (UB), Packed Benign (PB), Unpacked Malware (UM), and Packed Malware (PM). Let the ratio of data composition of dataset $D$ be $\alpha:\beta:\gamma:\delta$, where $\alpha$, $\beta$, $\gamma$, and $\delta$ are the proportions of Unpacked Benign (UB), Packed Benign (PB), Unpacked Malware (UM), and Packed Malware (PM) samples, respectively.

Based on the data composition above, the malware classifier $\pi$ is trained to achieve good accuracy using the labeled dataset $D$. The malware classifier $\pi$ along with the Dataset $D$ is fed as input to the TRUSTEE framework~\cite{jacobs2022ai} to produce an interpretable surrogate tree for the given black-box malware classifier model $\pi$. The TRUSTEE framework also provides importance scores for the features that contribute to the decision-making process for a sample to be classified as malware or benign. We consider the top $k$ features with non-zero importance scores, where $k$ might vary depending on the number of features exceeding the threshold in any iteration.
Let the top-feature set produced by TRUSTEE in the $t^{th}$ run be $S_t = \{ f_{t,1}, f_{t,2}, \ldots, f_{t,k} \}$, where each $f_{t,i}$ is a feature ranked as important during iteration $t$.

To obtain reliable and reproducible importance scores for the features, we repeated the above model training and TRUSTEE application $I$ times. The malware classifier $\pi$ is trained on the dataset $D$ generated with sampling constraints to reduce the overlap between data points in $I$ iterations, resulting in $I$ important feature sets. Let the important feature sets produced across the $I$ iterations be $S=\{ S_1, S_2, \ldots, S_I \}$, where each $S_t$ is a set of top-ranked features extracted in the $t^{th}$ iteration.

Note that as our framework is iterated for $I$ times, we obtain $I$ feature sets. Furthermore, this process ensures the influence of random initialization and sampling on model training and the TRUSTEE feature importance scores are less. The next step is to manually analyze and interpret the important features obtained, which is discussed in the next section.

\subsection{Manual Feature Analysis}
Now that we have the important feature sets $\mathcal{S}$ for the $I$ iterations, 
a strong manual analysis by the domain expert for interpreting the important features is required to understand 
the black-box malware classifier $\pi$ decisions and also the reason behind the 
importance of each feature. We propose analyzing each feature set $S_t$, 
where $t < I$, individually to determine whether $S_t$ is dominated by 
true binary behavior-related features. This is achieved by analyzing each feature
$f_{t,i}$, where $i = 1,\ldots,k$, within the feature set $S_t$ with non-zero 
importance score. Byte-level analysis of the representative PE files by the domain expert is performed 
by locating the presence of feature $f_{t,i}$ in PE files. Then, we inspect the surrounding 
bytes and the PE section in which $f_{t,i}$ appears to understand the context of 
the feature.  A frequency analysis was also performed to determine how many samples in each category (UB, PB, UM, PM) contained the given important feature $f_{t,i}$. This helps identify any patterns in the model's decisions and the interpretation of a particular feature. Finally, each feature $f_{t,i}$ was labeled as binary behavior-related or as an unintended artifact based on our analysis.

Now that we have the label for the feature $f_{t,i}$ we can determine if the feature set $S_t$ is binary behavior dominant or not, which in return tells us the performance of our malware classifier model $\pi$. The reliability of classifier $\pi$ is assessed by implementing the above framework under different dataset compositions to identify systematic biases in the features on which it relies. 

To evaluate the generality of the proposed framework, we extend the above methodology to additional classifiers and datasets. The same feature extraction and analysis pipeline is applied to models with different learning paradigms and to datasets with varying distributions of PE characteristics. This allows us to assess whether the identified feature dependencies are consistent across models and robust to dataset variations. Importantly, the methodology itself remains unchanged, as the TRUSTEE-based feature extraction and manual analysis are model-agnostic and can be applied to any black-box classifier trained on static features. This ensures that our framework is not tied to a specific model or dataset, but instead provides a general approach for analyzing feature dependencies in malware classification.
\section{Implementation}\label{sec:imp}

To evaluate the effectiveness of the proposed framework we aim to answer the following questions:\\
\textbf{RQ1:} Is the malware classifier learning true binary behavior or packing-related artifacts?\\  
\textbf{RQ2:} Are the identified feature dependencies consistent across models?  \\
\textbf{RQ3:} Do these dependencies persist across datasets?

\subsection{Experimental Setup}{\label{sec:exp}}
\begin{table*}[t]
\centering
\caption{Dataset composition counts and ratios.}
\label{tab:dataset}

\begin{subtable}{0.48\textwidth}
\centering
\caption{Dataset composition counts.}
\label{tab:counts_num}
\begin{footnotesize}
\begin{tabular}{|l|r|r|r|r|r|}
\hline
{Experiment} & {UB} & {PB} & {UM} & {PM} & {Total} \\
\hline
1. Biased dataset & 4396 & 0    & 0    & 4396 & 8792 \\ \hline
2. Mild benign packing introduction  & 3517 & 879  & 0    & 4396 & 8792 \\ \hline
3. Balanced benign samples & 2637 & 1758 & 0    & 4396 & 8792 \\ \hline
4. Benign packing dominant regime & 1758 & 2637 & 0    & 4396 & 8792 \\ \hline
5. High benign packing condition & 879  & 3517 & 0    & 4396 & 8792 \\ \hline
6. Fully packed setting & 0    & 4396 & 0    & 4396 & 8792 \\ \hline
\end{tabular}
\end{footnotesize}
\end{subtable}
\hfill
\begin{subtable}{0.48\textwidth}
\centering
\caption{Dataset composition ratios.}
\label{tab:counts_ratios}
\begin{footnotesize}
\begin{tabular}{|l|r|r|r|r|}
\hline
{Experiment} & $\alpha$ & $\beta$ & $\gamma$ & $\delta$ \\ \hline
1. Biased dataset & 0.50 & 0    & 0    & 0.50 \\ \hline
2. Mild benign packing introduction & 0.40 & 0.10 & 0    & 0.50 \\ \hline
3. Balanced benign samples & 0.30 & 0.20 & 0    & 0.50 \\ \hline
4. Benign packing dominant regime & 0.20 & 0.30 & 0    & 0.50 \\ \hline
5. High benign packing condition & 0.10 & 0.40 & 0    & 0.50 \\ \hline
6. Fully packed setting  & 0    & 0.50 & 0    & 0.50 \\ \hline
\end{tabular}
\end{footnotesize}
\end{subtable}

\end{table*}

To measure the effect of packing on windows PE files, we conducted several experiments following the architecture discussed in the previous section. As discussed in the Section~\ref{met1} each experiment was unique in terms of the data composition ratio $\alpha:\beta:\gamma:\delta$ as shown in Table~\ref{tab:counts_ratios}.

To evaluate the generality of the observed feature dependencies, we conducted additional experiments using an alternative classifier and a separate dataset with similar static-analysis features. The same top features set extraction and analysis pipeline was applied to ensure consistency. Standard evaluation metrics, including accuracy and F1-score, were used to confirm comparable model performance across all settings.

\subsubsection{Datasets}\label{sec:dataset}
The dataset used for all experiments in RQ1 and RQ2 was the \texttt{wild} dataset~\cite{aghakhani2020malware}, which is tabular data with 33{,}681 packed malicious, 12{,}647 packed benign, and 4{,}396 unpacked benign samples, for a total of 50{,}724 samples. We selected 8{,}792 samples for each experiment based on the experiment, which will be discussed in later sections.
The dataset has a dimensionality of 56{,}526 and consists of 9 families of static-analysis features. We removed a few features, such as \texttt{malicious\_vt} and \texttt{benign\_vt}, to limit the reliance on these features.

For RQ3 we used the EMBER dataset \cite{2018arXiv180404637A}, which provides a standardized feature representation of PE files, including byte histograms, byte-entropy distributions, strings, imports, exports, and header metadata.

\subsubsection{Malware Classifier Model}

We use TabNET model~\cite{arik2021tabnet} as our malware classifier model. TabNET is one of the best deep-architecture models for tabular datasets. TabNET uses sequential attention and feature selection at every decision step, selecting the most relevant features at every decision step. The ability of TabNET to handle large dimensional datasets by also providing high accuracies made us lean more towards TabNET as our malware classifier Model $\pi$ as the dataset $D$ is large.

The training accuracy was ensured to be at least greater than 90$\%$ for every five runs in the six experiments. The dataset was split in a ratio of 64:16:20 (Train:Validate:Test) to ensure stable training. TabNET was configured with 16 decision and attention steps $(n_d, n_a)$. The decision steps $n_{steps}$ was set to 3. Adam optimizer was used with a learning rate of 1e-3 and weight decay of 1e-5. To avoid overfitting, an early stopping method was applied, followed by the validation and testing phase, ensuring accuracies at least greater than 90$\%$. 

Additionally, we employ an XGBoost classifier~\cite{Chen_2016} to evaluate cross-model consistency and generalization. The model uses 500 gradient-boosted trees (max depth = 8, learning rate = 0.05) and is trained using a stratified 64:16:20 train/validation/test split. We ensured the accuracies of the classifiers for every experiment performed to answer RQ2 and RQ3 was above 95$\%$.

\subsubsection{Computing Environment}

All experiments were conducted on a workstation running Ubuntu~22.04.5~LTS. The system was equipped with a Intel Core i9-13900K CPU, 64\,GB DDR5 RAM along with NVIDIA GeForce RTX 4080 GPU with 16\,GB VRAM.
Two isolated Python environments, Python 3.10.16 was used training and for TRUSTEE application Python 3.13.2 was utilized.

\subsubsection{TRUSTEE configurations}

Post-hoc feature selection was performed using the TRUSTEE framework~\cite{jacobs2022ai} to understand the important features that contributed to the model's decision. In every experimental setup, the trained TabNET model with an accuracy greater than 90$\%$ along with the dataset, was fed as input to the TRUSTEE framework to obtain an interpretable decision tree and trust report. The number of inner loop iterations $N$ was set to 20 to generate $N$ student models. The inner loop parameter $S$ was set to 10 to select the best student model among  the 10 high-fidelity decision trees. The sample size was set to 0.1, which is 10$\%$ of the whole dataset fed as input for every iteration.  
\section{Evaluation}
\begin{figure*}[!t]
    \centering
    \includegraphics[width=0.7\linewidth]{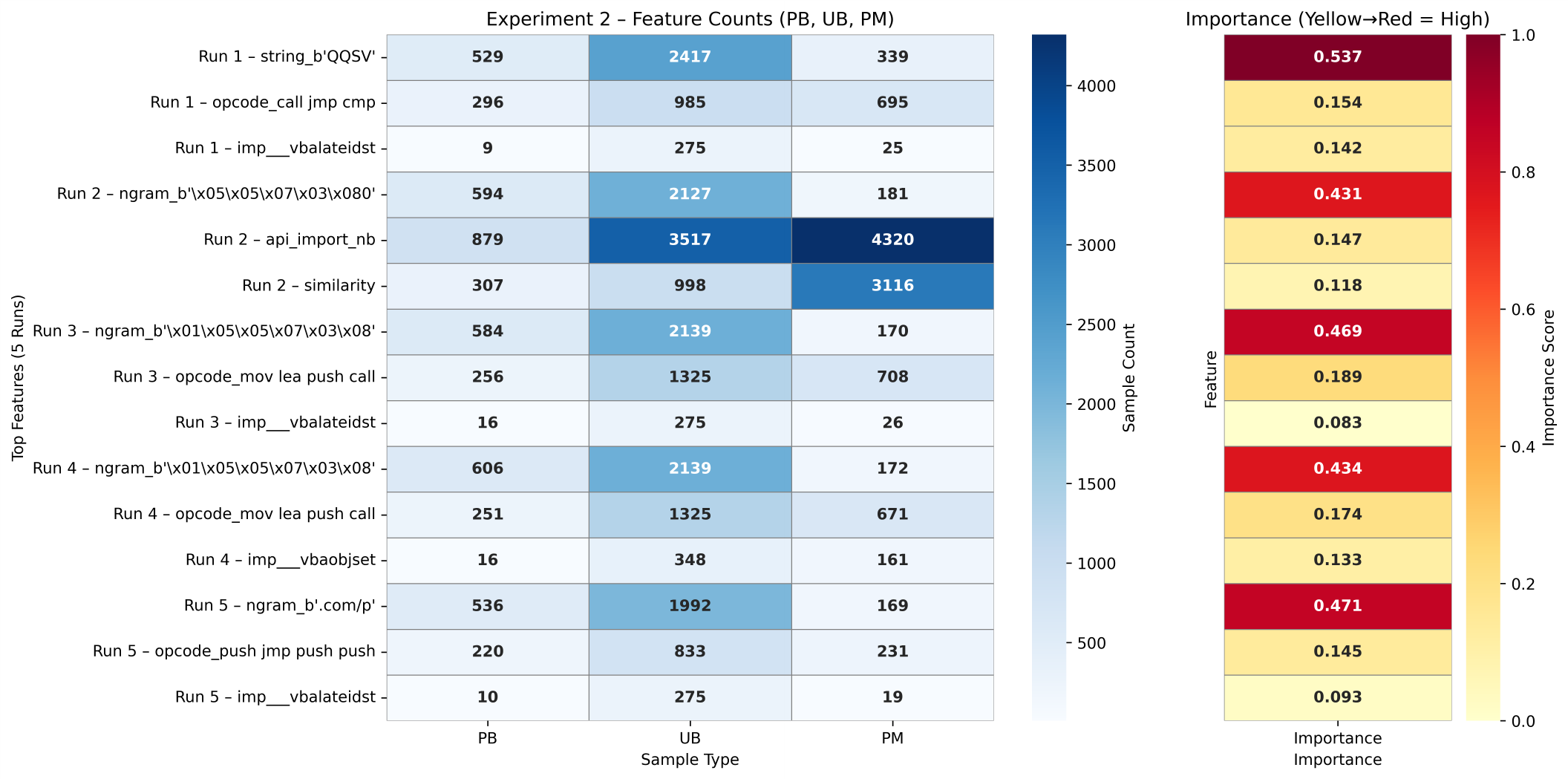}
    \caption{Top features presence and importance heatmaps for mild benign packing introduction experiment.}
    \label{fig:exp2}

\end{figure*}

\subsection{Feature Attribution under Controlled Data Compositions}
\begin{quote}
 \textbf{RQ1}: Is the malware classifier actually learning the true binary behavior or packing related artifacts?\   
\end{quote}

To answer the RQ1, we conducted six experiments with different data composition ratios  $(\alpha:\beta:\gamma:\delta)$ as shown in Table~\ref{tab:counts_ratios}. We chose I=5 for the proposed framework, thus creating five datasets that overlap minimally for every experiment, where each dataset has 8{,}792 samples. The TabNET models were trained with these datasets followed by TRUSTEE application and feature analysis was performed on the feature sets $S=\{ S_1, S_2, \ldots, S_I \}$. 
Table~\ref{tab:counts_ratios} summarizes the dataset composition used in all six experiments, showing the relative ratios and absolute sample counts for each class in Table~\ref{tab:counts_num}. Finally, the top-ranked features were manually analyzed in collaboration with cybersecurity experts to provide domain-informed interpretations of the model’s behavior.



\begin{table}[t]
{
\centering
\caption{Top three features across five runs for biased dataset experiment.}
\label{tab:exp1_top3}
\footnotesize
\renewcommand{\arraystretch}{1.1}
\setlength{\tabcolsep}{2pt}
\begin{tabular}{|l|l|r|r|r|r|}
\hline
{Run} & {Feature} & {\makecell{Impor-\\tance}} & \makecell{PB\\Count} & \makecell{UB\\Count} & \makecell{PM\\Count} \\
\hline

1 & \texttt{string\_b`QQSV'} & 0.527 & 0 & 3018 & 339 \\
  & \texttt{similarity} & 0.200 & 0 & 1234 & 3067 \\
  & \texttt{sectionsMaxEntropy} & 0.148 & 0 & 4396 & 4396 \\
\hline

2 & \texttt{string\_b`QQSV'} & 0.454 & 0 & 3018 & 373 \\
  & \texttt{similarity} & 0.165 & 0 & 1234 & 3116 \\
  & \texttt{sectionsMaxEntropy} & 0.162 & 0 & 4396 & 4396 \\
\hline

3 & \texttt{string\_b`QQSV'} & 0.426 & 0 & 3018 & 365 \\
  & \texttt{opcode\_jmp\_mov\_test} & 0.156 & 0 & 1454 & 800 \\
  & \texttt{imp\_\_vbalateidst} & 0.093 & 0 & 344 & 26 \\
\hline

4 & \texttt{opcode\_inc\_add\_inc\_add} & 0.525 & 0 & 3408 & 628 \\
  & \texttt{opcode\_jmp\_mov\_push\_push} & 0.097 & 0 & 1271 & 451 \\
  & \texttt{imp\_\_vbalateidst} & 0.085 & 0 & 344 & 21 \\
\hline

5 & \texttt{string\_b`QQSV'} & 0.505 & 0 & 3018 & 367 \\
  & \texttt{similarity} & 0.136 & 0 & 1234 & 3071 \\
  & \texttt{imp\_\_vbastrcat} & 0.112 & 0 & 426 & 160 \\
\hline

\end{tabular}%
}
\end{table}
\subsubsection{Biased to Balanced Feature Analysis}
In the biased dataset experiment, the dataset configuration $(0.5, 0, 0, 0.5)$ consisted solely of unpacked benign (UB) and packed malware (PM) samples as shown in Table~\ref{tab:counts_num}. Throughout all iterations, the most prominent feature was the string \texttt{b'QQSV'} as shown in Table~\ref{tab:exp1_top3}, which was frequently found in UB samples (at least 3{,}350 occurrences) but was mostly missing in PM samples \cite{almajed2025imbalance}. A byte-level examination revealed that this sequence is located in the \texttt{.text} section of unpacked binaries and corresponds to the x86 instruction pattern \texttt{push ecx; push ecx; push ebx; push esi}~\cite{intelISA,pietrek2002depth}. These binaries also consistently featured a \textit{Rich header}, a metadata structure specific to compilers associated with MSVC toolchains\cite{webster2017finding}, confirming that \texttt{QQSV} is a compiler artifact rather than an indicator of behavior. In contrast, PM samples lacked these patterns because packing either compressed or encrypted \cite{5633410} and instead showed high-entropy sections (e.g., \texttt{sectionsMaxEntropy}) indicative of compressed or encrypted content~\cite{lyda2007using}. This creates a distinct separation where the classifier $\pi$ differentiates classes using compiler artifacts versus entropy signals, rather than inherent malicious behavior.

In the second experiment which is mild benign packing introduction experiment, the introduction of packed benign samples $(0.40, 0.10, 0, 0.50)$ weakened the direct link between packing and maliciousness. In this scenario, byte $n$-grams became the dominant features as shown in Figure~\ref{fig:exp2}. The highest ranked $n$-gram were \texttt{ngram\_b`\textbackslash x01\textbackslash x05\textbackslash x05\textbackslash x07\textbackslash x03\textbackslash x08'}, \texttt{ngram\_b`\textbackslash x05\textbackslash x05\textbackslash x07\textbackslash x03\textbackslash x080'}.  To analyze these $n$-grams, we inspected the surrounding bytes in PE files that possess this n-gram and found that this feature lies in the overlay region in PE files, especially within authentic signature blocks along with certificate metadata, such as Certificate Revocation List distribution points(CRL), timestamp fields, and issuer metadata~\cite{microsoft_crypto_tools}. The byte patterns were encoded in ASN.1 (Abstract Syntax Notation .1), which is a binary encoding standard for digital signatures~\cite{microsoft_authenticode_timestamp}, thus making these $n$-grams certificate-related rather than behavioral features~\cite{kim2017certified}. Moreover, these features were more pronounced in the UB and PB samples than in the PM samples, as shown in the heatmap in Figure~\ref{fig:exp2}.

In the third experiment which is balanced benign samples $(0.30, 0.20, 0, 0.50)$, this trend continued. Certificate-related $n$-grams remained dominant, confirming ongoing reliance on signature metadata. Additionally, UTF-16 encoded substrings \cite{aghakhani2020malware} found in \texttt{.text}, \texttt{.rdata}, and \texttt{.rsrc} sections of PE files corresponding to API names and installer-related text emerged as secondary features as shown in Table~\ref{tab:exp3_top3}. These textual artifacts were more apparent in benign samples and less so in packed malware due to compression effects. Opcode-level features appeared only marginally and we interpret them as behavioral signals\cite{10.1007/978-3-319-50127-7_11}\cite{10.1016/j.ins.2011.08.020}, since they directly reflect executable code patterns in this study.
\begin{figure*}[!t]
    \centering
    \includegraphics[width=0.7\linewidth]{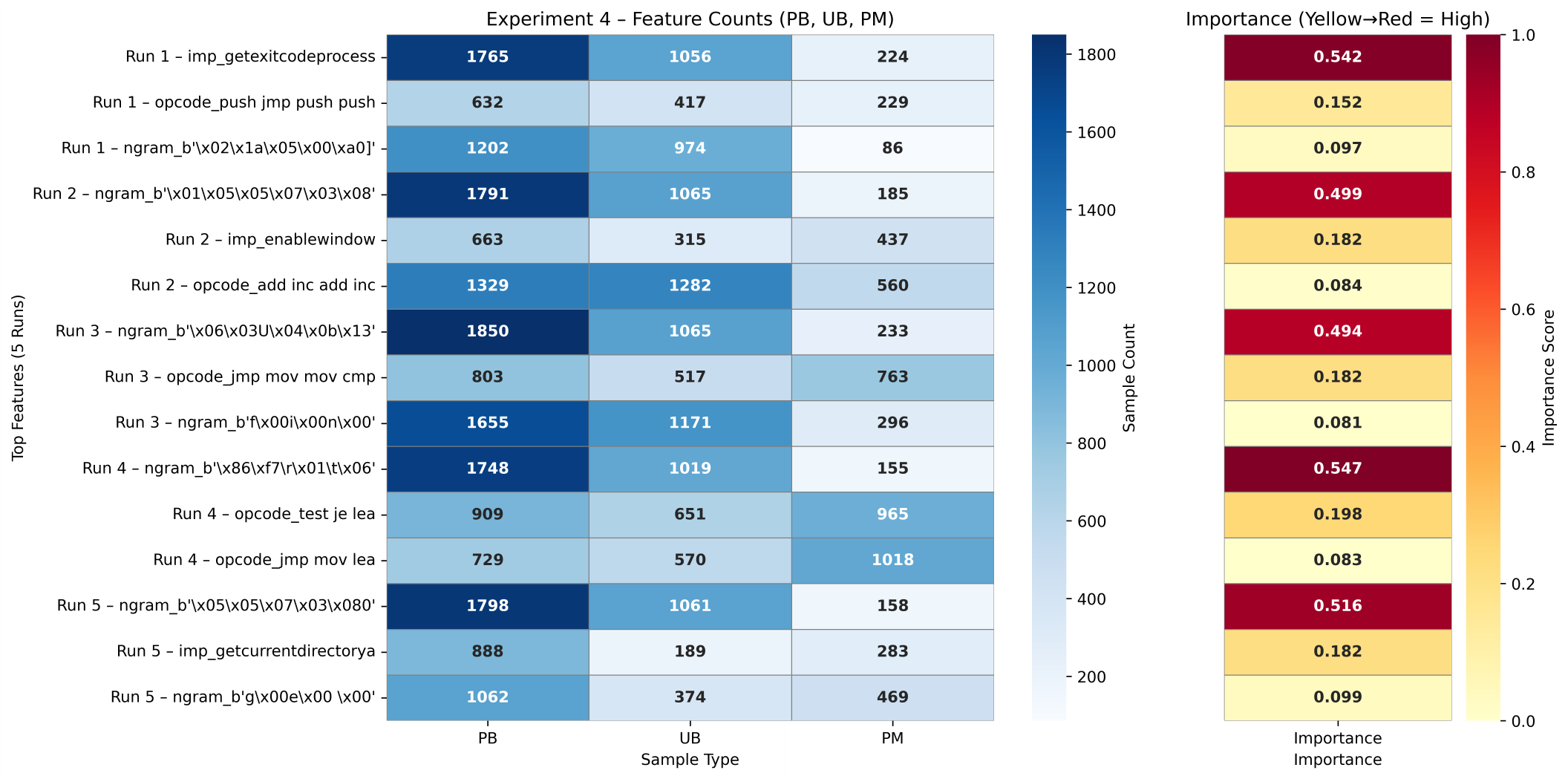}
    \caption{Top features presence and importance heatmaps for benign packing-dominant regime experiment.}
    \label{fig:exp4}
\end{figure*}
\begin{table}[t]
{
\centering
\caption{Top three features across five runs for balanced benign samples experiment.}
\label{tab:exp3_top3}
\footnotesize
\renewcommand{\arraystretch}{1.1}
\setlength{\tabcolsep}{1.2pt}

\resizebox{\columnwidth}{!}{%
\begin{tabular}{|l|l|r|r|r|r|}
\hline
{Run} & {Feature} & {\makecell{Impor-\\tance}} & \makecell{PB\\Count} & \makecell{UB\\Count} & \makecell{PM \\Count} \\
\hline

1 & \texttt{ngram\_b`\textbackslash x01\textbackslash x05\textbackslash x05\textbackslash x07\textbackslash x03\textbackslash x08'} 
  & 0.555 & 1174 & 1615 & 180 \\
  & \texttt{ngram\_b`\textbackslash x00a\textbackslash x00c\textbackslash x00t'} 
  & 0.187 & 801 & 573 & 455 \\
  & \texttt{ngram\_b`\textbackslash x08\textbackslash x00\textbackslash x09\textbackslash x00\textbackslash x0a\textbackslash x00'} 
  & 0.074 & 334 & 276 & 191 \\
\hline

2 & \texttt{ngram\_b`\textbackslash r\textbackslash x01\textbackslash x09\textbackslash x06\textbackslash x82'} 
  & 0.540 & 1171 & 1543 & 162 \\
  & \texttt{opcode\_push\_jmp\_push\_push} 
  & 0.148 & 413 & 625 & 281 \\
  & \texttt{imp\_\_vbalateidst} 
  & 0.110 & 18 & 201 & 17 \\
\hline

3 & \texttt{ngram\_b`\textbackslash x01\textbackslash x05\textbackslash x05\textbackslash x07\textbackslash x03\textbackslash x08'} 
  & 0.464 & 1201 & 1615 & 170 \\
  & \texttt{ngram\_b`l\textbackslash x00l\textbackslash x00a\textbackslash x00'} 
  & 0.174 & 719 & 790 & 270 \\
  & \texttt{opcode\_cmp\_jne\_cmp\_jne} 
  & 0.130 & 586 & 969 & 924 \\
\hline

4 & \texttt{ngram\_b`\textbackslash x05\textbackslash x05\textbackslash x07\textbackslash x03\textbackslash x08'} 
  & 0.459 & 1192 & 1607 & 167 \\
  & \texttt{opcode\_cmp\_je\_push\_call} 
  & 0.202 & 568 & 1022 & 695 \\
  & \texttt{ngram\_b`\textbackslash x00A\textbackslash x00n\textbackslash x00'} 
  & 0.077 & 591 & 1413 & 98 \\
\hline

5 & \texttt{ngram\_b`\textbackslash x06\textbackslash x03U\textbackslash x04\textbackslash x0b\textbackslash x13'} 
  & 0.482 & 1228 & 1616 & 220 \\
  & \texttt{ngram\_b`\textbackslash x00s\textbackslash x00t\textbackslash x00a'} 
  & 0.150 & 1361 & 2092 & 999 \\
  & \makecell[l]{\texttt{pesectionProcessed\_entrypoint} \\ \texttt{Sectioncharacteristics\_bit31}}
  & 0.092 & 147 & 2 & 1576 \\
\hline

\end{tabular}%
}
}

\end{table}

\subsubsection{Packing Dominant to Fully Packed Feature Analysis}
In the experiment benign packing dominant regime $(0.20, 0.30, 0, 0.50)$, the proportion of packed benign (PB) samples was further increased, resulting in a setting dominated by benign packing. Despite this change, the most significant features continued to be certificate-related byte $n$-grams, which appeared frequently in PB samples (approximately 1{,}748--1{,}850 occurrences) and much less in PM samples (150--250 occurrences) as shown in Figure~\ref{fig:exp4}. UTF-16 encoded string patterns also ranked highly, indicating embedded textual artifacts within PE sections. Opcode sequences (e.g., \texttt{opcode\_jmp mov lea}) and import-based features (e.g., \texttt{imp\_enablewindow}) were of moderate importance, likely due to the more balanced sample composition\cite{235493}. However, these signals remained secondary and did not strongly differentiate the samples. Overall, the classifier $\pi$ primarily relied on certificate- and string-based artifacts rather than behavioral characteristics.

\begin{table}[!htbp]
\centering
\caption{Top three features across five runs for high benign packing condition experiment.}
\label{tab:exp5_top3}
\small
\setlength{\tabcolsep}{4pt}
\renewcommand{\arraystretch}{1.15}
\resizebox{\columnwidth}{!}{%
\begin{tabular}{|l|l|r|r|r|r|}
\hline
{Run} & {Feature} & {\makecell{Impor-\\tance}} & {\makecell{PB\\Count}} & {\makecell{UB\\Count}} & {\makecell{PM\\Count}} \\
\hline
1 & \texttt{ngram\_b`://www'}                                            & 0.558 & 2598 & 579  & 403  \\
  & \texttt{ngram\_b`S\textbackslash 00T\textbackslash 00U\textbackslash 00'} & 0.078 & 1319 & 346  & 132  \\
  & \texttt{opcode\_jmp\_mov\_cmp}                                       & 0.149 & 1075 & 235  & 707  \\
\hline
2 & \texttt{ngram\_b`\textbackslash x86\textbackslash f7\textbackslash r\textbackslash x01\textbackslash t\textbackslash x05'} & 0.570 & 2425 & 548 & 244 \\
  & \texttt{opcode\_jmp\_mov\_push\_push}                                & 0.157 & 999  & 221  & 466  \\
  & \texttt{opcode\_add\_pop\_mov}                                       & 0.085 & 678  & 197  & 1127 \\
\hline
3 & \texttt{ngram\_b`\textbackslash x06\textbackslash x03U\textbackslash x04\textbackslash x0b\textbackslash x13'} & 0.575 & 2448 & 549 & 233 \\
  & \texttt{opcode\_push\_mov\_xor}                                      & 0.127 & 1251 & 306  & 1015 \\
  & \texttt{opcode\_mov\_sar}                                            & 0.069 & 804  & 213  & 1078 \\
\hline
4 & \texttt{ngram\_b`\textbackslash r\textbackslash x01\textbackslash x01\textbackslash x05\textbackslash x05\textbackslash x00'} & 0.572 & 2413 & 534 & 219 \\
  & \texttt{opcode\_test\_je\_lea}                                       & 0.192 & 1223 & 305  & 965  \\
  & \texttt{ngram\_b`\textbackslash xff\textbackslash xff\textbackslash x85\textbackslash xc0\textbackslash x10'} & 0.087 & 1994 & 403 & 152 \\
\hline
5 & \texttt{ngram\_b`\textbackslash x0b\textbackslash x06\textbackslash t*\textbackslash x86H'} & 0.619 & 2353 & 536 & 155 \\
  & \texttt{opcode\_jmp\_mov\_push\_push}                                & 0.169 & 1040 & 221  & 468  \\
  & \texttt{ngram\_b`f\textbackslash 00i\textbackslash 00n\textbackslash 00'} & 0.061 & 2141 & 617 & 325 \\
\hline
\end{tabular}%
}
\end{table}
In high benign packing condition $(0.10, 0.40, 0, 0.50)$ experiment, this pattern persisted. Certificate-related $n$-grams again dominated feature importance, including patterns linked to certificate overlays and metadata fields \cite{kim2017certified} as shown in Table~\ref{tab:exp5_top3}. Additional UTF-16 substrings which are neighbors to the features discussed in previous experiments appeared and these features are related to resource and version information with moderate importance. As in previous settings, opcode and import-based features contributed only marginally, reinforcing that classification decisions were driven by metadata and encoding patterns rather than executable behavior.

\begin{figure*}[!t]
    \centering
    \includegraphics[width=0.7\linewidth]{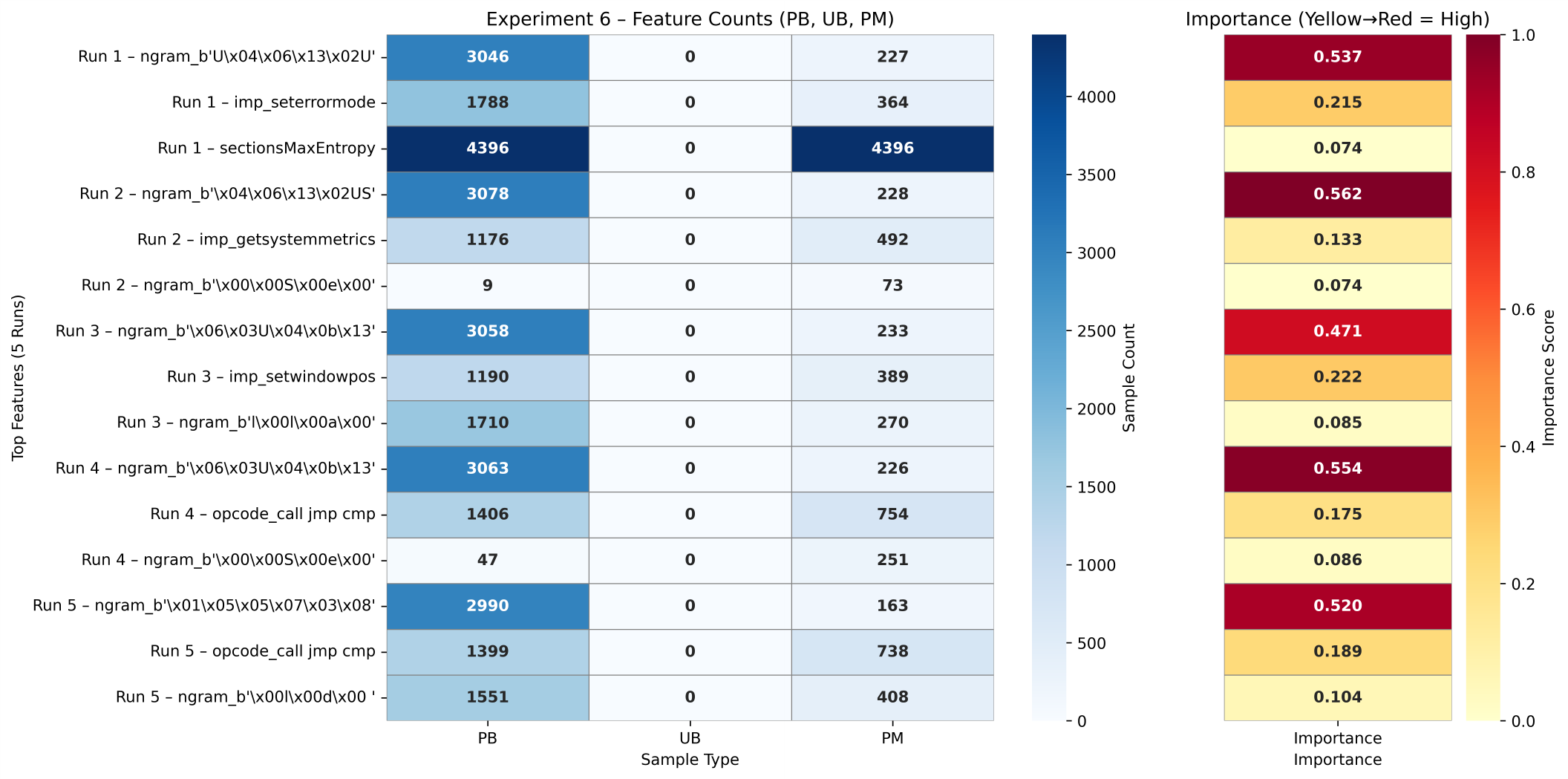}
    \caption{Top features presence and importance heatmaps for fully packed setting experiment.}
    \label{fig:exp6}
\end{figure*}

In fully packed setting $(0, 0.50, 0, 0.50)$ experiment, a fully packed setting was examined where both benign and malicious samples were uniformly packed. Even with the complete removal of unpacked samples, the most influential features remained certificate-related $n$-grams and string-based patterns as shown in Figure~\ref{fig:exp6}. These features were mainly associated with benign samples, while malicious samples lacked similar certificate structures, allowing the classifier to continue distinguishing classes based on metadata presence.

\begin{tcolorbox}[colback=blue!5!white,colframe=black!75!black,boxsep=1pt,left=3pt,right=3pt,top=1pt,bottom=1pt, title=Findings \#1, label=box:Findings1]
Across every dataset composition examined in this study, the malware classifier was depending on packing artifacts, certificate blocks or resource level metadata rather than learning true binary behavior.
\end{tcolorbox}
\begin{table}[!htbp]
\centering
\caption{Families of $n$-grams.}
\label{tab:ngram_families}
\small
\renewcommand{\arraystretch}{1.1}

\resizebox{\columnwidth}{!}{
\begin{tabular}{|l|r|}
\hline
{$n$-gram Family} & {$n$-grams} \\ \hline

Certificate-related 
& 
\makecell[r]{
\texttt{ngram\_b`U\textbackslash x04\textbackslash x06\textbackslash x13\textbackslash x02U'} \\
\texttt{ngram\_b`\textbackslash x04\textbackslash x06\textbackslash x13\textbackslash x02US'} \\
\texttt{ngram\_b`\textbackslash x06\textbackslash x03U\textbackslash x04\textbackslash x0b\textbackslash x13'} \\
\texttt{ngram\_b`\textbackslash x01\textbackslash x05\textbackslash x05\textbackslash x07\textbackslash x03\textbackslash x08'}
}
\\ \hline

UTF-16 related
&
\makecell[r]{
\texttt{ngram\_b`\textbackslash x00o\textbackslash x00d\textbackslash x00u'} \\
\texttt{ngram\_b`\textbackslash x00\textbackslash x20\textbackslash x00F\textbackslash x00a'} \\
\texttt{ngram\_b`l\textbackslash x00l\textbackslash x00a\textbackslash x00'}
}
\\ \hline

\end{tabular}%
}

\end{table}
\subsection{Cross-Model Consistency of Artifact Dependencies}
\begin{quote}\textbf{RQ2}: Are the identified feature dependencies consistent across different models?\end{quote}
\subsubsection{Biased to Balanced Feature Analysis}

\begin{table}[t]
\centering
\caption{Top feature and importance across five runs for cross-model consistency from biased dataset to balanced benign samples.}
\label{tab:exp1_3_top1}
\renewcommand{\arraystretch}{1.2}
\resizebox{\columnwidth}{!}{
\begin{tabular}{|l|l|r|}
\hline
{Experiment} & {Top Feature (Runs 1--5)} & {Importance} \\
\hline
{Biased dataset}
& \texttt{string\_b`QQSV'}              & 0.527 \\
& \texttt{string\_b`QQSV'}              & 0.454 \\
& \texttt{string\_b`QQSV'}              & 0.426 \\
& \texttt{opcode\_inc\_add\_inc\_add}   & 0.525 \\
& \texttt{string\_b`QQSV'}              & 0.505 \\
\hline
{\makecell[l]{Mild benign\\packing introduction}}
& \texttt{string\_b`QQSV'}                                                  & 0.494 \\
& \texttt{ngram\_b`oratio'}                                                 & 0.509 \\
& \texttt{ngram\_b`\textbackslash x01\textbackslash x05\textbackslash x05\textbackslash x07\textbackslash x03\textbackslash x08'} & 0.520 \\
& \texttt{ngram\_b` Code '}                                                 & 0.522 \\
& \texttt{ngram\_b`\textbackslash x01\textbackslash x05\textbackslash x05\textbackslash x07\textbackslash x03\textbackslash x08'} & 0.477 \\
\hline

{\makecell[l]{Balanced\\benign samples}}
& \texttt{ngram\_b`\textbackslash x01\textbackslash x05\textbackslash x05\textbackslash x07\textbackslash x03\textbackslash x08'} & 0.504 \\
& \texttt{ngram\_b`U\textbackslash x04\textbackslash x06\textbackslash x13\textbackslash x02U'}                                   & 0.586 \\
& \texttt{ngram\_b`\textbackslash x05\textbackslash x05\textbackslash x07\textbackslash x03\textbackslash x080'}                  & 0.545 \\
& \texttt{ngram\_b`\textbackslash x01\textbackslash x05\textbackslash x05\textbackslash x07\textbackslash x03\textbackslash x08'} & 0.548 \\
& \texttt{ngram\_b`\textbackslash x1c\textbackslash x06\textbackslash t*\textbackslash x86H'}                                     & 0.544 \\
\hline
\end{tabular}%
}
\end{table}
To determine whether the artifact-driven behavior identified in RQ1 generalizes across models, we trained XGBoost classifier~\cite{Chen_2016} with 500 trees (max depth = 8, learning rate = 0.05) using a stratified train/validation/test split on the same datasets used in RQ1 experiments. Rather than re-analyzing individual features, the focus was on whether similar dependency patterns emerged. The results closely mirror those observed in RQ1. In biased dataset experiment, XGBoost again prioritized compiler-related artifacts such as \texttt{QQSV}, along with string- and entropy-based features, confirming its reliance on visibility differences introduced by packing as shown in Table~\ref{tab:exp1_3_top1}. In the Mild benign packing introduction and balanced benign samples experiments, the model shifted toward certificate-related $n$-grams and UTF-16 textual fragments, consistent with the behavior of $\pi$. Although feature importance appeared slightly more distributed, the dominant signals remained unchanged byte-level patterns, metadata, and structural properties while opcode and behavior based features played a secondary role. Overall, these experiments indicate that the artifact reliance observed in RQ1 is not model-specific. XGBoost captures the same correlations, suggesting that these dependencies stem from dataset characteristics rather than the learning algorithm.

\begin{table}[t]
\centering
\caption{Top feature and importance across five runs for cross model consistency from benign packing dominant regime to fully packed setting.}
\label{tab:exp4_6_top1}
\mytablesize
\renewcommand{\arraystretch}{1.1}

\resizebox{\columnwidth}{!}{
\scalebox{0.75}{%
\begin{tabular}{|l|l|r|}
\hline
{Experiment} & {Top Feature (Runs 1--5)} & {Importance} \\
\hline

\makecell[l]{Benign packing \\dominant regime} &
\makecell[l]{
\texttt{ngram\_b`\textbackslash x06\textbackslash x03U\textbackslash x04\textbackslash x0b\textbackslash x13'} \\
\texttt{ngram\_b`\textbackslash x06\textbackslash x13\textbackslash x02US1'} \\
\texttt{ngram\_b`\textbackslash x01\textbackslash x05\textbackslash x05\textbackslash x07\textbackslash x03\textbackslash x08'} \\
\texttt{ngram\_b`\textbackslash x06\textbackslash x03U\textbackslash x04\textbackslash x0b\textbackslash x13'} \\
\texttt{ngram\_b`\textbackslash x06\textbackslash x03U\textbackslash x04\textbackslash x0b\textbackslash x13'}
}
&
\makecell[r]{
0.511 \\
0.509 \\
0.548 \\
0.495 \\
0.503
}
\\
\hline

\makecell[l]{High benign \\packing condition} &
\makecell[l]{
\texttt{ngram\_b`\textbackslash x86\textbackslash xf7\textbackslash x0d\textbackslash x01\textbackslash x09\textbackslash x06'} \\
\texttt{ngram\_b`\textbackslash x05\textbackslash x05\textbackslash x07\textbackslash x03\textbackslash x080'} \\
\texttt{ngram\_b`U\textbackslash x04\textbackslash x06\textbackslash x13\textbackslash x02U'} \\
\texttt{ngram\_b`\textbackslash x06\textbackslash x13\textbackslash x02US1'} \\
\texttt{ngram\_b`\textbackslash x01\textbackslash x05\textbackslash x05\textbackslash x07\textbackslash x03\textbackslash x08'}
}
&
\makecell[r]{
0.523 \\
0.570 \\
0.528 \\
0.489 \\
0.560
}
\\
\hline

\makecell[l]{Fully packed \\ setting} &
\makecell[l]{
\texttt{ngram\_b`\textbackslash x05\textbackslash x05\textbackslash x07\textbackslash x03\textbackslash x080'} \\
\texttt{ngram\_b`\textbackslash x06\textbackslash x03U\textbackslash x04\textbackslash x0b\textbackslash x13'} \\
\texttt{ngram\_b`\textbackslash r\textbackslash x01\textbackslash x09\textbackslash x061\textbackslash x82'} \\
\texttt{ngram\_b`\textbackslash x06\textbackslash x13\textbackslash x02US1'} \\
\texttt{ngram\_b`\textbackslash x05\textbackslash x05\textbackslash x07\textbackslash x03\textbackslash x080'}
}
&
\makecell[r]{
0.593 \\
0.470 \\
0.572 \\
0.635 \\
0.528
}
\\
\hline

\end{tabular}%
}}

\end{table}

\subsubsection{Packing Dominant to Fully Packed Feature Analysis}
A similar trend is observed in packing dominant regime to fully packed setting experiments. Consistent with RQ1, byte-level $n$-grams particularly those associated with certificates remain the most influential features across all configurations as shown in Table~\ref{tab:exp4_6_top1}. Structural features such as entropy and section statistics appear more consistently, while opcode and API-related features contribute moderately due to the more balanced data composition. However, these higher-level signals remain secondary to metadata-driven artifacts. In the fully packed setting, the model exhibits an even stronger concentration on a small subset of highly discriminative $n$-grams, further emphasizing its dependence on dataset-specific encoding patterns rather than behavioral semantics.

\begin{tcolorbox}[colback=blue!5!white,colframe=black!75!black,boxsep=1pt,left=3pt,right=3pt,top=1pt,bottom=1pt, title=Findings \#2, label=box:Findings2]
RQ2 confirms that the behavior identified in RQ1 transfers directly across models. While XGBoost incorporates a slightly broader range of features, its decision-making remains fundamentally driven by the same non-semantic artifacts, reinforcing the conclusion that these patterns are intrinsic to the dataset rather than dependent on model choice.
\end{tcolorbox}
\subsection{Cross-Dataset Generalization}
\begin{quote}\textbf{RQ3:} \textit{Do these feature dependencies persist across different datasets?}\end{quote}

\begin{table}[t]
\centering
\caption{Top features and their importance scores in cross-dataset generalization experiments.}
\label{tab:rq3_top_features}
\small
\begin{tabular}{|l|l|r|}
\hline
{Run} & {Feature Name} & {Importance} \\
\hline
1 & \texttt{byteentropy\_77} & 1.000000 \\
\hline
2 & \texttt{imports\_USER32.dll\_18=GetDC} & 1.000000 \\
\hline
3 & \texttt{strings\_urls} & 0.774859 \\
\hline
3 & \texttt{histogram\_131} & 0.225141 \\
\hline
\end{tabular}
\end{table}
While RQ2 demonstrated that artifact driven feature dependencies remain consistent across different models, it remains unclear whether these patterns are specific to the \textit{wild} dataset\cite{aghakhani2020malware} or indicative of a broader phenomenon. To investigate this, we evaluated whether similar behavior arises under a different data distribution.

To this end, we conducted experiments using the EMBER dataset \cite{2018arXiv180404637A}, which provides a standardized feature representation of PE files, including byte histograms, byte-entropy distributions, strings, imports, exports, and header metadata. Using an XGBoost classifier and the same feature attribution framework, we analyzed which feature categories dominate model predictions in this setting.

The results reveal a consistent and striking pattern: in each run, a single feature overwhelmingly dominates with an importance score close to $1.0$ as shown in Table~\ref{tab:rq3_top_features}, while all other features contribute negligibly\cite{perasso2025empirical}. Although the dominant feature varies across runs, it consistently belongs to a small subset of feature groups, such as byte entropy features, import-based features, and string-level features. Byte entropy features capture local randomness and are strongly associated with packing or compression; import features reflect API-level interactions; and string features encode embedded textual content such as URLs and metadata. Despite spanning different semantic categories, the model relies almost exclusively on one feature at a time, rather than combining multiple complementary signals.

The classifier depends on a single highly discriminative feature instead of learning a robust combination of structural, behavioral, and semantic characteristics. Notably, this observation is consistent with the findings from RQ2, where the model similarly concentrated on narrow feature groups such as $n$-grams and metadata artifacts.
\begin{tcolorbox}
[colback=blue!5!white,colframe=black!75!black,boxsep=1pt,left=3pt,right=3pt,top=1pt,bottom=1pt, title=Findings \#3, label=box:Findings3]
Overall, the EMBER results suggest that this behavior is not confined to a specific dataset or model configuration, but rather reflects a broader tendency of malware classifier models to prioritize easily separable statistical cues over deeper program semantics\cite{perasso2025empirical}. This has important implications for generalization, as such models are likely to fail under distribution shifts or adversarial transformations that disrupt these dominant features.
\end{tcolorbox}
\section{Conclusion}
This study investigated whether machine learning based malware classifiers capture intrinsic binary behavior or instead rely on non-semantic artifacts. To examine this question, we conducted controlled experiments under varying dataset compositions and applied TRUSTEE-based feature attribution complemented by manual byte-level analysis.

Findings from RQ1 show that the classifier consistently prioritizes non-semantic features, including compiler artifacts, entropy-based packing indicators, certificate metadata, and embedded strings. Even when dataset composition is adjusted to weaken simple correlations, the model adapts by exploiting alternative artifact-driven signals rather than learning genuine behavioral characteristics such as instruction semantics or control-flow structures.

RQ2 demonstrates that this behavior is not model-specific. When evaluated with XGBoost, similar dependency patterns emerge, with the model relying on the same categories of artifact-driven features despite differences in learning mechanisms. This suggests that the observed behavior is primarily shaped by dataset properties rather than the model architecture.

RQ3 further establishes that these dependencies generalize across datasets. Experiments on the EMBER dataset reveals that the classifier relies almost entirely on a single highly discriminative feature (e.g., entropy, imports, or strings) instead of combining multiple complementary signals. This indicates that dependence on superficial statistical cues is a broader phenomenon, not confined to a particular dataset.

Overall, these findings demonstrate that high classification accuracy does not necessarily imply learning of true malicious behavior. Instead, the models evaluated predominantly exploit dataset-specific artifacts with little emphasis on behavioral signals raising concerns about their robustness and generalization under distribution shifts. While similar patterns are observed across both models and datasets considered in this study, suggesting a strong influence of dataset characteristics, further investigation across additional models and data distributions is required.

\noindent\textbf{(a) Implications for Robust Malware Detection:}
These results suggest that improving robustness in malware classification requires moving beyond artifact-driven learning. This includes: (i) constructing datasets that explicitly control for biases introduced by packing, compilers, and metadata; (ii) including more behavioral signals by adding dynamic analysis features; and (iii) regularization or interpretability-guided training is encouraged which increases the dependence on multiple feature types.

\noindent\textbf{(b) Future Work:}
Future work will focus on integrating dynamic behavioral features, designing artifact-invariant learning objectives, and developing robustness metrics that explicitly penalize reliance on non-semantic features. This study underscores the importance of carefully balanced datasets, robust interpretability techniques, and domain expertise in the evaluation of malware classifiers.

\bibliographystyle{IEEEtran}
\bibliography{bibliography}
\end{document}